\documentclass[twocolumn,twoside]{article}
\usepackage{graphics}
\usepackage[T1,T2A]{fontenc}  
\usepackage[utf8]{inputenc}
\usepackage[ukrainian,english]{babel}

\newcommand{\hb}{\\ \hspace*{2ex}}

\begin{document}

\title{NUMERICAL ESTIMATOR FOR LARGE-SCALE COSMIC STRUCTURES}
\author{A.\,V. Tugay$^{1}$, N.\,G. Pulatova$^{2}$, A.\,D. Zhoga$^{1}$ \\[2mm] 
\begin{tabular}{l}
 $^1$ Taras Shevchenko National University of Kyiv,\hb
 Glushkova 4, Kyiv, 03022, Ukraine,  {\em tugay.anatoliy@gmail.com}\\
 $^2$ Main Astronomical Observatory, \hb
 Zabolotnogo, 27, Kyiv, 03143, Ukraine, {\em nadya@mao.kiev.ua}\\[2mm]
\end{tabular}
}
\date{}
\maketitle
ABSTRACT. 
Components of large-scale structure (LSS) of Universe includes galaxy clusters, walls, filaments, groups and field galaxies. 
The question of spatial organization of all these components remains open despite of a number of recently developed methods of LSS analysis. 
In this paper we introduce new continuous structural parameter of galaxy distribution for the determinig of the type of LSS surrounding. 
This parameter is based on comparison of the radial distribution of galaxies around the certain point with the uniform one. 
Our method for structure parameter calculation is based on the distribution of 1000 closets galaxies to the selected one. 
Such number was selected because most galaxy clusters contains up to 1000 galaxies. 
When we exceed 1000th closest galaxy we will leave possible cluster or filament and will pass to another type of LSS. 
We suppose that in uniform distribution of galaxies their number should increase as third power a distance, similar to luminosity. 
Therefore the inverse dependence of galaxy distance from number should be cubic root. 
New structural parameter was introduced as the difference of this dependence of cubic root. 

The main type of cosmic surrounding for a galaxy is defined by the value and the sign of such structural parameter.
If a galaxy lie in a cluster the distances of nearby galaxies will be less than cubic root. 
We integrated distance difference for all 1000 galaxies and get negative stuctural parameter in this case. 
If a galaxy lie in a void, the distances of surrounding galaxies will be larger than cubic root. In such case the structural parameter will be positive.  

New parameter was calculated for sky distribution of SDSS galaxies at distances from 50 to 150 Mpc. 
Analysing obtained values, we come to conclusion that our structural parameter can be used for the division of LSS components and extragalactic filament detection. \\[1mm]
{\bf Keywords}: large-scale structure of Universe, galaxies, filaments.\\[2mm]
КІЛЬКІСНИЙ ПАРАМЕТР ВЕЛИКОМАСШТАБНИХ КОСМІЧНИХ СТРУКУТУР.
Компоненти великомасштабної структури Всесвіту (ВМС) включають скупчення галактик, стіни, філаменти, групи і галактики поля. Питання про просторове розташування всіх цих компонент залишається відкритим незважаючи на велику кількість сучасних методів аналізу ВМС. У даній роботі ми вводимо новий неперервний структурний параметр розподілу галактик для визначення типу оточення у ВМС. Визначення параметру полягає у порівнянні радіального розподілу галактик навколо заданої точки з однорідним розподілом. При обчисленні структурного параметра для обраної галактики використовується розподіл тисячі найближчих до неї галактик. Таке число було вибране оскільки більшість скупчень містять до тисячі галактик. Після вичерпання тисячі найближчих галатик можливе скупчення або філамент має закінчитися і ми перейдемо до інших елементів ВМС. Ми припускаємо, що у випадку однорідного розподілу кількість галактик (а також їх повна світність) має зростати пропорційно кубу відстані. Отже, обернена залежність відстані галактики від її порядкового номера має бути близька до кубічного кореня. Новий структурний параметр вводиться як інтеграл відхилення цієї залежності від кубічного кореня. 

Тип великомасштабного оточення галактики визначається величиною і знаком структурного параметра. Якщо галактика знаходиться у скупченні, відстані сусідніх галактик будуть менші, ніж за усередненою залежністю з кубічним коренем. При додаванні різниць двох залежностей для тисячі галактик отримується від'ємний структурний параметр. Якщо ж галактика знаходиться у войді, відстані сусідніх галактик будуть більші за кубічний корінь з номера. У цьому випадку структурний параметр буде додатним.

Новий параметр був обчислений для розподілу по небу галактик SDSS на відстанях від 50 до 150 Мпк. Аналіз отриманих значень дозволяє зробити висновок про те, що такий параметр може бути використаний для розділення компонентів ВМС і виявлення міжгалактичних філаментів.\\[1mm]
{\bf Ключові слова}: великомасштабна структура Всесвіту, галактики, філаменти.
\\[2mm]
%
%
{\bf 1. Introduction}\\[1mm]

Large-scale structure (LSS) of Universe is complex but mostly covered from direct optical observations. Galaxies are the only type of light emitting sources at cosmological scale, but they posess minor part of total mass of matter. 
In some special cases it is possible to find the distribution of dark matter in galaxy clusters by pecular velocities or gravitational lensing (Kaiser, 1993; Parnovsky, 2004).
There are different views on distribution of LSS elements. 
Results of analysis of observations and simulations are controversal. 
The most common view on LSS is to consider voids bordered by walls, walls consisting from filaments and filaments which has clusters, groups and field galaxies. In this work we propose a method for numerical evaluating the type of LSS element for any point of extragalactic space based on the coordinates of nearby galaxies. The method was tested with modern catalogs of isolated galaxies (Karachentseva, 2010; Karachentsev, 2011).\\[2mm]

{\bf 2. Types of LSS elements and methods of their description}\\[1mm]

In the first case of uniform distribution of galaxies in space we will consider the luminosity of galaxies as the main observable parameter. For physical analysis of LSS evolution the mass is more important, but it can be evaluated correct for very little part of galaxies. The study of mass/luminosity ratio is wide and complex direction of extragalactic astronomy (Girardy, 2000) but in this work we will explain LSS only by the distribution of visible light of galaxies.
Moreover, in the current paper we will assume that all galaxies has the same luminosity and hence the large-scale cosmic structures will be analysed only by galaxy positions and density. If all galaxies are indentical and little enough, their total number and luminosity will increase as the third power of radial distance. Such uniform distribution should be teseted for observable Universe at the scales much less and larger than 100 Mpc, the main scale of cosmic web. 
If the number of galaxies around the chosen point grow as the first power of distance, nearby LSS should be considered as one-dimensional, i.e. filament. If luminosity increase as squared radius, such region will be called a wall.

To determine the morphological type of local LSS we build the cumulative luminosity function (CLF) for the chosen point. As it was mentioned before, we count only the number of galaxies but not magnitudes and M/L relation for CLF estimation. An excess or depression could be found for CLF at lower scales. If there is an excess we can say that chosen point (or central galaxy) lies in cluster. If there is a depression, then the galaxy should be located in the void. 
Now we introduce the large-scale structural parameter as normalized depression at cumulative luminosity function. It should have largest values for isolated galaxies and low (negative) values for cluster galaxies. The main cases of possible usage of structural parameter are the following.

1. One LSS element in uniform matter distribution. We can find its type by local power index of CLF and the value of the structural parameter. The shape of CLS is important for detailes description of current structure. 
For example, it is possible to detect galaxy cluster and to select appropriate radial profile of dark matter distribution for it. For detailed description of each LSS element a number of numerical structural parameters are needed, but even two parameters can not be calculated for real data of galaxy observations.

2. Complex system of LSS elements repeating periodically in the space. Suppose that void has radius R and walls are thin. Then we should not build CLF for distances larger than 2R. We can expect large number of observale galaxies at this range for the Local Universe. By the CLF shape it is possible to describe LSS in details. But in the ultimate worst case there could be no galaxies at all. Such opposite situations lead to the following issues. Is was supposed above that all galaxies are similar. But if we will consider the brightest galaxies (for periodic LSS), each of them will be the central galaxy of a cluster at the intersection of some number of voids, walls and filaments. For example, in graphite-like spatial grate each cluster is the point of connection of 4 voids, 6 walls and 4 filaments. 
The number of mentioned brightest galaxies should be proportional to the number of voids. 
For the current state of observational astronomy it is easy to find a distance at which even one observed galaxy per void or cluster can not be guaranteed. 
This distance is a limit of any assumption of uniform distribution of galaxy positions and luminosities.

3. Discrete distribution of different single galaxies behind the mentioned limit. That galaxies are very bright and tends to have random distribution. Is it possible calculate their structural parameters and put them to some place in LSS? Consider a galaxy, its nearest neighbour and the luminosiites of these two. It is important to measure the minimal distance between galaxies in the units of void radius. This parameter can be used for the description of distant Universe unstead of CLF shape. The brightest galaxies in the Universe will appead once per many LSS cells. Very probably they will be beamed AGNs and this is only one from a number of selection effects at such large distances. 
In this extreme case of many cells between two uniquely bright galaxies we can fill space between them with normal distribution of faint galaxies with some upper luminosity limit. It will be some kind of model of distant galaxy distribution with apriori parameters. In the rest part of a paper we will consider parameters of galaxy distribution from much closer region.

In this paper we describe LSS by special structural parameter anticorrelated with density contrast of dark matter mass and galaxy luminosity. 
Consider a list of cosmic structures in the order of increasing isolation.

1. Central galaxy of a cluster

2. Galaxy near the cluster center

3. Galaxy in the periphery of cluster

4. Galaxy in a group

5. Galaxy in a pair or triplet

6. Field galaxy in the filament

7. Isolated galaxy in the wall

8. Isolated galaxy in the void

We suppose here that LSS consist from cells with walls at the borders and voids inside the cells. Walls themselves consist from filaments. Groups and clusters may appear in the filaments according to necklace model. The largest clusters tends to be found at the intersection of filaments. In this work we will consider the following LSS elements:

1. Cluster. Galaxy is counted as galaxy in cluster if it can pass cluster diameter at Hubble time. Galaxies in cluster are tied by common gravity field. So groups, triplets, pairs and interacting galaxies are considered as clusters. Cluster is compact overdensity in spatial galaxy distribution. It can be surrounded by ellipsoid with some limit ellipticity.

2. Filament. It is one-dimensional galaxy overdensity. Filament can be curved but we can not describe the shape of filament with our structural parameter.
According to inconsistency of different methods for filament detection from observational data, the locations and forms of certain real filaments seems to be doubtful. Structural parameter, introduced in the present work, is the next attempt to distinguish filaments from visible galaxy distribution.

3. Wall. Although filaments may not so obvious, there are well defined underdensities in the Universe i.e. voids. The volume between voids is much larger than the volume of clusters. This volume is generally referenced as walls. It is natural to consider the walls as two-dimensional overdensities.

4. Voids should contain isolated and field galaxies but not clusters and groups. There is a problem to locate the border between void and wall. 
For example, in the case of two nearby maximums in the distribution of structural parameter, then they should be assigned as voids and the space between them should include the wall.

These elements describe a medium in which the structural parameter is calculated. Its value increases from cluster to void. 
The method of normalization of our structural parameter is the following. To reduce the effect of peculiar velocities in redshift space, we take a sample of galaxies in 100 Mpc redshift span and don't use their distances at the next stages.
This corresponds to suggestion that all galaxies of our sample lie in a single layer of LSS. 
Hence all galaxies are placed on celestial sphere with the radius equal to unity. 
Then the normalized distances were found from one selected galaxy to another N=1000 nearby galaxies. Such distances (R) are measured in angular units i.e. radians. 
The averaged angular distance of nearby galaxies is an extra value for local LSS description. 
The value of the structural parameter is an integrated difference of cubic root function from R(N) dependence. 
So this value is the excess of average angular distance above uniform distribution.\\[2mm]

{\bf 3. Distribution of structural parameter for real galaxies}\\[1mm]

To estimate optimal distance range for CLF and structural parameter calculation we found the following bounds.
 Hyperleda extragalactic database contains 18619 galaxies within v3k=3500 km/s that corresponds Local Supecluster. We illustrate the completeness of current observational data with D/L relation, where D is average distance between nearst observable galaxies and L=100 Mpc - the size of cosmic web cell.
 At z=0.2 (6 layers of superclusters) the ratio D/L=0.2 and at z=1 D/L=4. 
 There are 3 layers and 180 voids at low redshift (z<0.1) with 1000 galaxies per void. According such relations we concluded that the method of CLF analysis is correct at distances up to 1000 Mpc. 

In this work we selected from HyperLeda database redshift region from 3000 to 10000 km/s that corresponds to Coma supercluster position. 
It is the most appropriate sample of galaxies for LSS study (Tugay, 2012; Tugay, 2014). 

We find three main types of R(N) dependences. 

1. Cluster-like galaxy surrounding. The most galaxies has distances less than cubic root and structural parameter is negative. An example of such dependence is shown in a Fig. 1. 

2. Void-like surrounding. The most galaxies has distances larger than cubic root and structural parameter is positive (Fig. 2). 
 
3. Intermediate case in which N(R) dependence intersect cubic root one or more times (Fig. 3). In such case structural parameter may lie in wide range of positive and negative values. 
The distribution of number of intersections from structural parameter is shown at Fig. 4. 
The value of structural parameter is the difference of galaxy angular distance (in radians) from cubic root function averaged for one thousand nearby galaxies. 
To distinguish void-like regions we selected the values of structural parameter less than -0.03. Negative values from -0.03 to 0 were considered as corresponding to filament, wall or group regions.

Galaxies in void-like surrounding has much larger distances to neighbours. We analysed the distances to 1000th closest galaxy as secondary structural parameter (SSP). To estimate how CLF characterise isolated galaxies, SSP was also calculated for 2MASS selected isolated galaxy catalog 2MIG (Karachentseva, 2010). Results are presented in Table 1. SSP is systematically larger for 2MIG galaxies. 

Sky distributions of galaxies with different values of structural parameter are presented at Fig 5-7. These distributions leads to the following conclusions.

1. Structural parameter is smoothe continous value. Largest clusters, such as Coma at Fig. 7, can not be detected with it. 

2. Cluster-like regions represents systematical effects of Hyperleda sky coverage and has minor correlation with real clusters.

3. The most clear picture of cosmic web can be seen only in the zone of SDSS coverage (120<RA<240 deg, 0<DEC<60 deg), where observational data are the most dense and uniform. Regions of intermediate values of structural parameter could be used for filament detection.\\[2mm]

\begin{table}[h]
\caption{Secondary structural parameter for HyperLeda and 2MIG galaxies measured in centiradians (34 arcmin). SSP is angular distance to 1000th nearest galaxy. HL-2MIG is HyperLeda sample without 3227 galaxies with largest SSP. Only minor part of these galaxies are 2MIG members.}
\begin{tabular}{lrrr}
\hline
Sample          & HyperLeda & 2MIG    & HL-2MIG \\
\hline
Num. of galaxies& 89904     & 3227    & 86677   \\
Minimal SP      &  1.1      &  4.5    &  1.1    \\
Average SP      & 11.1496   & 16.8932 & 10.9358 \\
Error of avg.SP &  0.0161   &  0.1171 &  0.0159 \\
Scatter         &  4.8      &  6.7    &  5.0    \\
Maximal SP      & 38.1      & 38.1    & 22.4    \\
\hline
\end{tabular}
\end{table}

{\bf 4. Conclusion}\\[1mm]

We developed a new method for LSS description by the distribution galaxy positions. 
New structural parameter can be used for defining LSS elements at distances up to 1000 Mpc. \\[2mm]

{\bf References\\[2mm]}
Girardi M., Borgani S., Giuricin G. et al. 2000. {\it ApJ,} {\bf 530}, 62 \\
Kaiser N., Squires G. 1993. {\it ApJ,} {\bf 404}, 441 \\	
Karachentsev I.D., Makarov D.I., Karachentseva V.E., Melnyk O.V. 2011. {\it Astrophysical Bulletin,} {\bf 66}, 1 \\ 
Karachentseva~V.\,E., Mitronova~S.\,N., Melnyk~O.\,V. \& Karachentsev~I.\,D. 2010. {\it ASPC,} {\bf 421}, 11 \\ 
Parnovsky S.L., Tugay A.V. 2004. {\it AstL,} {\bf 30}, 357 \\
Tugay A.V. 2012. {\it Odessa Astron. Publ.,} {\bf 25}, 142 \\
Tugay A.V. 2014. {\it Advances in Astronomy and Space Physics,} {\bf 4}, 42 \\

\begin{figure}[h] 
\resizebox{1.0\hsize}{!}
{\includegraphics{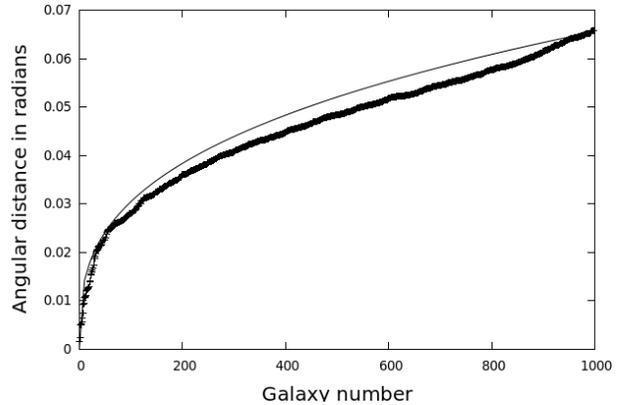}}
\label{fig1}
\caption{Cumulative luminosity function for a galaxy in cluster-like surrounding.}
\end{figure}

\begin{figure}[h] 
\resizebox{1.0\hsize}{!}
{\includegraphics{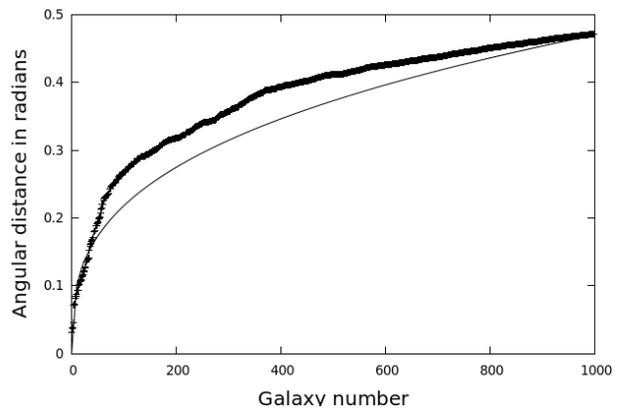}}
\label{fig2}
\caption{CLF for a galaxy in void-like surrounding.}
\end{figure}

\begin{figure}[h] 
\resizebox{1.0\hsize}{!}
{\includegraphics{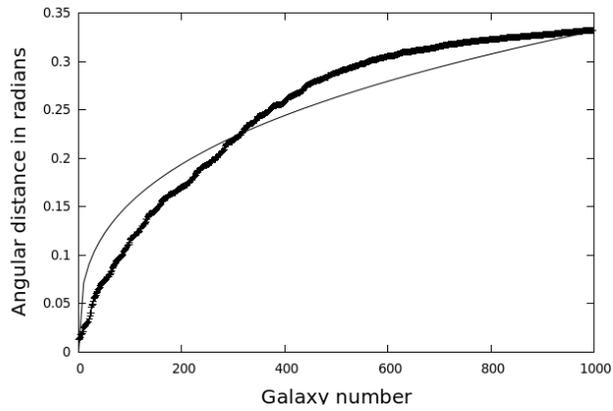}}
\label{fig3}
\caption{The intersection of uniform relation by CLF should be referred as wall or filament.}
\end{figure}

\begin{figure}[h] 
\resizebox{1.0\hsize}{!}
{\includegraphics{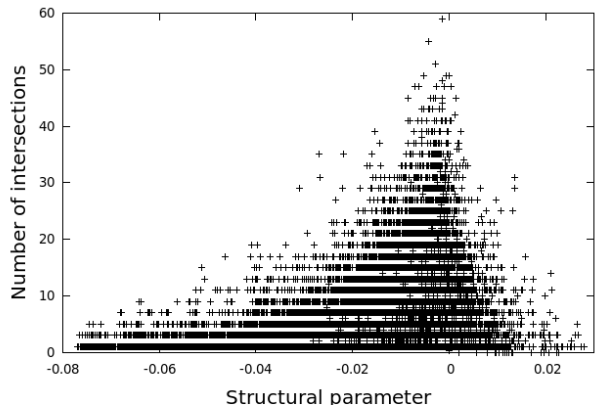}}
\label{fig4}
\caption{Distribution of number of intersections from structural parameter.}
\end{figure}

\begin{figure}[h] 
\resizebox{1.0\hsize}{!}
{\includegraphics{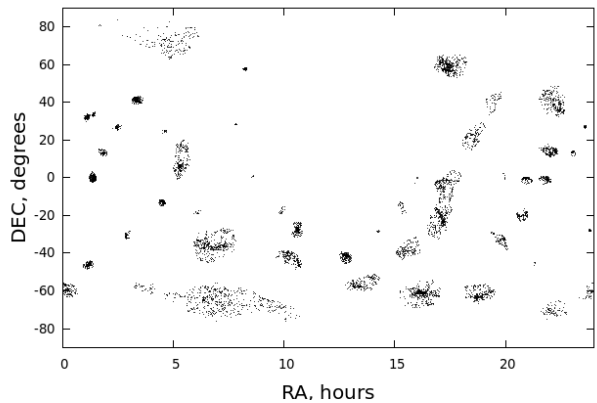}}
\label{fig5}
\caption{Sky distribution of HyperLeda galaxies with lowest value of structural parameters.}
\end{figure}

\begin{figure}[h] 
\resizebox{1.0\hsize}{!}
{\includegraphics{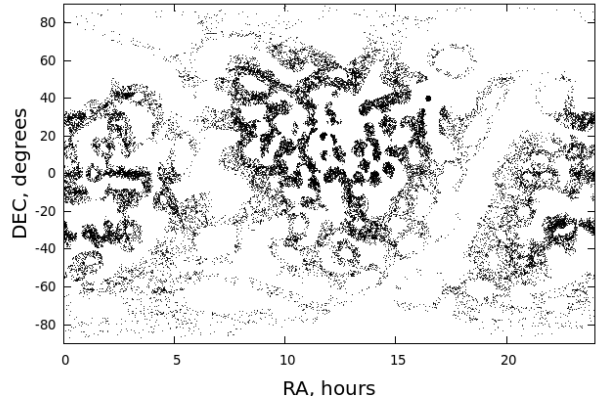}}
\label{fig6}
\caption{Regions with possible filaments.}
\end{figure}

\begin{figure}[h] 
\resizebox{1.0\hsize}{!}
{\includegraphics{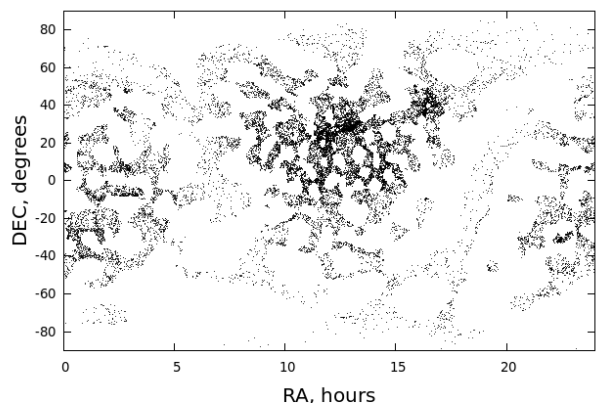}}
\label{fig7}
\caption{Void candidates according to largest values of structural parameter.}
\end{figure}

\vfill
\end{document}